\documentclass[aps,prd,superscriptaddress,twocolumn]{revtex4}
\usepackage[dvips]{graphicx}
\usepackage{bm,latexsym,amsmath,amssymb,amsfonts,mathrsfs}
\usepackage{color}
\input{colordvi.tex}
\newcommand*{\D}{{\rm d}}
\newcommand*{\mpl}{M_{\rm Pl}}

\begin{document}

\title{Primordial non-Gaussianities of gravitational waves in the most general
single-field inflation model}

\author{Xian~Gao}
\email[Email: ]{xgao@apc.univ-paris7.fr}
\affiliation{%
Astroparticule et Cosmologie (APC), UMR 7164-CNRS,
Universit\'{e} Denis Diderot-Paris 7, 10 rue Alice Domon et L\'{e}onie Duquet,
75205 Paris, France
}
\affiliation{%
Laboratoire de Physique Th\'{e}orique, \'{E}cole Normale Sup\'{e}rieure,
24 rue Lhomond, 75231 Paris, France
}
\affiliation{%
Institut d'Astrophysique de Paris (IAP), UMR 7095-CNRS,
Universit\'{e} Pierre et Marie Curie-Paris 6, 98bis Boulevard Arago, 75014 Paris, France
}

\author{Tsutomu~Kobayashi}
\email[Email: ]{tsutomu"at"tap.scphys.kyoto-u.ac.jp}
\affiliation{Hakubi Center, Kyoto University, Kyoto 606-8302, Japan
}
\affiliation{Department of Physics, Kyoto University, Kyoto 606-8502, Japan
}

\author{Masahide~Yamaguchi}
\email[Email: ]{gucci"at"phys.titech.ac.jp}
\affiliation{Department of Physics, Tokyo Institute of Technology, Tokyo 152-8551, Japan}

\author{Jun'ichi~Yokoyama}
\email[Email: ]{yokoyama"at"resceu.s.u-tokyo.ac.jp}
\affiliation{Research Center for the Early Universe (RESCEU), Graduate School of Science,
The University of Tokyo, Tokyo 113-0033, Japan}
\affiliation{Institute for the Physics and Mathematics of the Universe (IPMU),
The University of Tokyo, Kashiwa, Chiba, 277-8568, Japan}

\begin{abstract}
We completely clarify the feature of primordial non-Gaussianities of
tensor perturbations in generalized G-inflation, {\em i.e.,} the most
general single-field inflation model with second order field
equations. It is shown that the most general cubic action for the tensor
perturbation (gravitational wave) $h_{ij}$ is composed only of two
contributions, one with two spacial derivatives and the other with one
time derivative on each $h_{ij}$.  The former is essentially identical
to the cubic term that appears in Einstein gravity and predicts a
squeezed shape, while the latter newly appears in the presence of the
kinetic coupling to the Einstein tensor and predicts an equilateral
shape. Thus, only two shapes appear in the graviton bispectrum of the
most general single-field inflation model, which could open a new clue
to the identification of inflationary gravitational waves in
observations of cosmic microwave background anisotropies as well as
direct gravitational wave detection experiments.
\end{abstract}

\maketitle

Inflation, an accelerated expansion of the early Universe caused by a
scalar field called inflaton, is a quite promising paradigm of
cosmology, and primordial perturbations generated from inflation are
crucial clues to the yet unidentified inflationary model. From the
properties of the primordial perturbations such as the power spectra and
spectral indices, we can extract information about the theory governing
the inflaton dynamics.  Among them the non-Gaussian signature in the
cosmic microwave background (CMB) has been paid much attention in recent
years, along with the great progress in precise cosmological
observations.  So far most of the literature has focused upon
non-Gaussianities of the scalar perturbations~\cite{malda1}, as they are most directly
connected to the CMB observations. Tensor perturbations~\cite{Sta}, however, are
also generated during inflation, whose direct detection would be the
most obvious evidence for inflation. When we try to detect tensor
perturbations with the CMB measurements and/or with the direct detection
experiments, it is essentially important to remove the background
(contamination) sources. For example, the B-mode polarizations are
dominated by the lensing effects on relatively small
scales. Astrophysical sources like white dwarf binaries could dominate
the power spectrum for a wide range of frequencies of the
background gravitational waves. Thus, non-Gaussianities will be a
key feature of the tensor perturbations~\cite{malda2, soda} as well as the scalar
perturbations because they can help us to discriminate the inflationary
signals from other contamination sources even if the latter dominates
the power spectrum. For this purpose, we need to completely clarify the
features of the non-Gaussianities of primordial tensor perturbations
produced during inflation, which enables us to make templates for
non-Gaussianities of primordial gravitational waves.

In this {\em Letter}, we, for the first time, investigate the
non-Gaussianities of primordial tensor perturbations based on the most general
single-field inflation model, {\em i.e.}, generalized G-inflation~\cite{G2},
make a complete identification of the shapes of
bispectra, and explore the
possibility of large non-Gaussianities from the tensor sector.

The Lagrangian for generalized G-inflation is the most general one that
is composed of the metric $g_{\mu\nu}$, the scalar field $\phi$, and
their arbitrary derivatives, and has the second-order field equations.
The Lagrangian was first derived by Horndeski in 1974~\cite{Horndeski},
and very recently it was rediscovered in a modern form as the
generalized Galileon~\cite{GG}, {\em i.e.}, the most general extension
of the Galileon~\cite{Galileon, CovGali}, in four dimensions.  The
generalized Galileon is described by the sum of the following four:
\begin{eqnarray}
&&{\cal L}_2 = K(\phi, X),
\quad
{\cal L}_3=-G_3(\phi, X)\Box\phi,
\cr
&&{\cal L}_4=G_4(\phi, X)R+G_{4X}\left[(\Box\phi)^2-(\nabla_\mu\nabla_\nu\phi)^2\right],
\cr
&&{\cal L}_5=G_5(\phi, X)G_{\mu\nu}\nabla^\mu\nabla^\nu\phi
-\frac{1}{6}G_{5X}\bigl[(\Box\phi)^3
\cr
&&\qquad\qquad\qquad
-3\Box\phi(\nabla_\mu\nabla_\nu\phi)^2+
2(\nabla_\mu\nabla_\nu\phi)^3\bigr],
\end{eqnarray}
where $K$ and $G_i$'s are arbitrary functions of $\phi$ and $X:=-(\partial\phi)^2/2$.
Here we used the notation $G_{iX}$ for $\partial G_i/\partial X$.
The generalized Galileon can be used as a framework
to study the most general single-field inflation model.
Generalized G-inflation contains novel models, as well as
previously known models of single-field inflation such as standard canonical inflation,
k-inflation~\cite{kinf}, extended inflation~\cite{extinf},
$R^2$ inflation~\cite{R2inf}, new Higgs inflation~\cite{newhiggs},
and (minimal) G-inflation~\cite{Ginf}.
The above four Lagrangians can even reproduce the
non-minimal coupling to the Gauss-Bonnet term~\cite{G2}.

In~\cite{G2}, the background equations for generalized G-inflation is presented,
and the most general quadratic actions for tensor and scalar perturbations
are determined, giving the power spectra of the primordial perturbations.
The most general cubic action for scalar perturbations
is worked out in~\cite{Gao, Tsujikawa}.
The curvature perturbation in generalized G-inflation is shown to be conserved
on large scales at non-linear order in~\cite{Gao2}.
We are going to present the most general cubic action
for tensor perturbations to determine the possible tensor bispectrum
arising from single-field inflation.

The perturbed metric around a cosmological background may be written as
\begin{eqnarray}
g_{00}=-1,\quad g_{0i}=0,\quad g_{ij}=a^2(t) \left(e^h\right)_{ij},\label{metrich}
\end{eqnarray}
where
\begin{eqnarray}
\left(e^h\right)_{ij}=\delta_{ij}+h_{ij}+\frac{1}{2}h_{ik}h_{kj}+\frac{1}{6}h_{ik}h_{kl}h_{lj}+
\cdots,
\end{eqnarray}
and $h_{ij}$ is a transverse and traceless tensor perturbation,
$\partial_ih_{ij}=0 =h_{ii}$,
with repeated spatial indices summarized by $\delta_{ij}$.
The perturbed metric defined in this way is convenient for
calculating the action because we have $\sqrt{-g} = a^3$.
We plug the metric~(\ref{metrich}) into the action
\begin{eqnarray}
S=\sum_{i=2}^5\int\D^4 x\sqrt{-g}{\cal L}_i,
\end{eqnarray}
and expand it in terms of $h_{ij}$ to get the quadratic and cubic actions.
Only the Lagrangians that involve the curvature tensors and $\nabla_\mu\nabla_\nu\phi$,
{\em i.e.},
${\cal L}_4$ and ${\cal L}_5$,
contribute to the quadratic and cubic actions.

The quadratic action was already derived in~\cite{G2}:
\begin{eqnarray}
S^{(2)} =\frac{1}{8}\int\D t\D^3x\,a^3\left[
{\cal G}_T\dot h_{ij}^2-\frac{{\cal F}_T}{a^2}
(\partial_k h_{ij})^2\right], \label{tensoraction}
\end{eqnarray}
where an overdot denotes $\D/\D t$ and
\begin{eqnarray}
{\cal F}_T&:=&2\left[G_4
-X\left( \ddot\phi G_{5X}+G_{5\phi}\right)\right],
\\
{\cal G}_T&:=&2\left[G_4-2 XG_{4X}
-X\left(H\dot\phi G_{5X} -G_{5\phi}\right)\right].\;\;
\end{eqnarray}
From the action we see that
the propagation speed of the gravitational waves is
given by $c_h:=\sqrt{{\cal F}_T/{\cal G}_T}$,
which may differ from unity. In order for the system to be stable
${\cal F}_T>0$ and ${\cal G}_T>0$ are required.

The linear perturbation equation derived from the action~(\ref{tensoraction})
can be solved in the Fourier space,
\begin{eqnarray}
h_{ij}(t, \mathbf{x})=\int\frac{\D^3 k}{(2\pi)^3}\,
\tilde h_{ij}(t,\mathbf{k})e^{i\mathbf{k}\cdot\mathbf{x}}.
\end{eqnarray}
To proceed further, it is convenient to introduce
a new time coordinate defined by $\D y=c_h \D t/a$.
We assume for simplicity that the inflationary Universe may be
approximated by de Sitter spacetime and
${\cal F}_T$ and ${\cal G}_T$ are approximately constant.
Using the normalized mode solution,
\begin{eqnarray}
\psi_\mathbf{k} = \frac{\sqrt{\pi}}{a}
\sqrt{\frac{c_h}{{\cal F}_T}}
\sqrt{-y}H_{3/2}^{(1)}(-ky),
\end{eqnarray}
where $H^{(1)}_{3/2}$ is the Hankel function,
the quantized tensor perturbation is written as
\begin{eqnarray}
\tilde h_{ij}=\sum_s\left[\psi_{\mathbf{k}}e_{ij}^{(s)}(\mathbf{k})a_s(\mathbf{k})
+\psi^*_{-\mathbf{k}}e_{ij}^{*(s)}(-\mathbf{k})a_s^\dagger(-\mathbf{k})\right],
\end{eqnarray}
where $e_{ij}^{(s)}$ is the polarization tensor with the helicity states $s=\pm 2$,
satisfying $e_{ii}^{(s)}(\mathbf{k})=0=k_je_{ij}^{(s)}(\mathbf{k})$.
Here we adopt the normalization such that
$e_{ij}^{(s)}(\mathbf{k})e_{ij}^{*(s')}(\mathbf{k})=\delta_{ss'}$.
Choosing the phase of the polarization tensors appropriately, we have the relations
$e_{ij}^{*(s)}(\mathbf{k})=e_{ij}^{(-s)}(\mathbf{k})=e_{ij}^{(s)}(-\mathbf{k})$.
The commutation relation for the creation and annihilation operators is given by
$
[a_s(\mathbf{k}), a_{s'}^\dagger(\mathbf{k}')]=(2\pi)^3\delta_{ss'}\delta(\mathbf{k}-\mathbf{k'})
$.
The 2-point function can now be computed as
\begin{eqnarray}
\langle \tilde h_{ij}(\mathbf{k})\tilde h_{kl}(\mathbf{k}')\rangle
&=&
(2\pi)^3\delta^{(3)}(\mathbf{k}+\mathbf{k}'){\cal P}_{ij,kl}(\mathbf{k}),
\\
{\cal P}_{ij,kl}&=&
|\psi_{\mathbf{k}}|^2\Pi_{ij,kl}(\mathbf{k}),
\end{eqnarray}
where we introduced
\begin{eqnarray}
\Pi_{ij,kl}(\mathbf{k})=\sum_{s}
e_{ij}^{(s)}(\mathbf{k})e_{kl}^{*(s)}(\mathbf{k}).
\end{eqnarray}
The power spectrum ${\cal P}_h=(k^3/2\pi^2){\cal P}_{ij,ij}$
is given by
\begin{eqnarray}
{\cal P}_h(k)= \left.\frac{1}{2\pi^2}\frac{H^2}{{\cal F}_Tc_h}\right|_{k y=-1},
\end{eqnarray}
where
$y=-1/k$ corresponds to the time of the sound horizon exit.


Having thus obtained the quadratic action and
the solution to the linearized equation governing the inflationary gravitational waves,
we now move on to the cubic action.
The most general cubic action for tensor perturbations in the single-field context
is obtained as
\begin{eqnarray}
S^{(3)}&=&\int\D t\D^3x\,a^3\biggl[
\frac{X\dot \phi G_{5X}}{12} \dot h_{ij}\dot h_{jk}\dot h_{ki}
\nonumber\\&&\qquad
+\frac{{\cal F}_T}{4a^2}\left(h_{ik}h_{jl}
-\frac{1}{2}h_{ij}h_{kl}\right)\partial_k\partial_lh_{ij}
\biggr],
\end{eqnarray}
which is composed only of two contributions.
Clearly, the term with one time derivative on each $h_{ij}$
appears only if $G_{5X}\neq 0$.
This term is absent in the case of Einstein gravity,
non-minimal coupling to gravity (${\cal L}_3=f(\phi)R$),
and even in the case of new Higgs inflation which involves
a non-standard kinetic term of the form
$G^{\mu\nu}\partial_\mu\phi\partial_\nu\phi$ \footnote{This corresponds to
$G_5=-\phi$, and hence $G_{5X}=0$ in this case.}.
However, in the presence of non-minimal coupling to the Gauss-Bonnet term,
this term does not vanish~\cite{G2}.
The terms of the form $h^2\partial^2h$, where $\partial$ represents
a spatial derivative,
is already present
in the case of Einstein gravity,
and even in the most general case
only the overall normalization is
generalized from the Planck mass squared
$\mpl^2$ to the function ${\cal F}_T$.

The 3-point function can be computed by employing
the in-in formalism,
\begin{eqnarray*}
&&\langle \tilde h_{i_1j_1}(\mathbf{k}_1)
\tilde h_{i_2j_2}(\mathbf{k}_2)
\tilde h_{i_3j_3}(\mathbf{k}_3)
\rangle
\nonumber\\&&
=-i\int^t_{t_0}\D t'\langle [
\tilde h_{i_1j_1}(t, \mathbf{k}_1)
\tilde h_{i_2j_2}(t, \mathbf{k}_2)
\tilde h_{i_3j_3}(t, \mathbf{k}_3), H_{\rm int}(t')]\rangle,
\end{eqnarray*}
where $t_0$ is some early time when the perturbation is
well inside the sound horizon, $t$ is a time several e-foldings after
the sound horizon exit, and the interaction Hamiltonian is
\begin{eqnarray}
H_{\rm int}(t)=-\int\D^3x\,a^3\biggl[
\frac{X\dot \phi G_{5X}}{12} \dot h_{ij}\dot h_{jk}\dot h_{ki}+\cdots\biggr].
\end{eqnarray}
It will be convenient to introduce the non-Gaussian amplitude
${\cal A}_{i_1j_1i_2j_2i_3j_3}$ defined by
\begin{eqnarray}
&&\langle \tilde h_{i_1j_1}(\mathbf{k}_1)
\tilde h_{i_2j_2}(\mathbf{k}_2)
\tilde h_{i_3j_3}(\mathbf{k}_3)
\rangle
\nonumber\\&&\quad
=(2\pi)^7\delta^{(3)}(\mathbf{k}_1+\mathbf{k}_2+\mathbf{k}_3)
{\cal P}_h^2\frac{{\cal A}_{i_1j_1i_2j_2i_3j_3}}{k_1^3k_2^3k_3^3}.\label{def-amp}
\end{eqnarray}
We write
${\cal A}_{i_1j_1i_2j_2i_3j_3}={\cal A}^{({\rm new})}_{i_1j_1i_2j_2i_3j_3}+{\cal A}^{({\rm GR})}_{i_1j_1i_2j_2i_3j_3}$,
where ${\cal A}^{({\rm new})}_{i_1j_1i_2j_2i_3j_3}$
and ${\cal A}^{({\rm GR})}_{i_1j_1i_2j_2i_3j_3}$ represent the contributions from
the $\dot h^3$ term and the $h^2\partial^2 h$ terms, respectively.
Just for simplicity, here
again the inflationary Universe is approximated by
de Sitter spacetime,
which allows us to to compute
the non-Gaussian amplitude
assuming that
$X\dot\phi G_{5X}\simeq$ const and ${\cal F}_T\simeq$ const.
Each contribution is then found to be
\begin{widetext}
\begin{eqnarray}
{\cal A}_{i_1j_1i_2j_2i_3j_3}^{({\rm new})} &=&
\frac{HX\dot\phi G_{5X}}{4{\cal G}_T}
\frac{k_1^2k_2^2k_3^2}{K^3}\Pi_{i_1j_1, lm}(\mathbf{k}_1)
\Pi_{i_2j_2, mn}(\mathbf{k}_2)
\Pi_{i_3j_3, nl}(\mathbf{k}_3),
\\
{\cal A}_{i_1j_1i_2j_2i_3j_3}^{({\rm GR})} &=&
\tilde{\cal A}
\left\{
\Pi_{i_1j_1,ik}(\mathbf{k}_1)\Pi_{i_2j_2,jl}(\mathbf{k}_2)
\left[
k_{3k}k_{3l}\Pi_{i_3j_3,ij}(\mathbf{k}_3)
-\frac{1}{2}k_{3i}k_{3k}\Pi_{i_3j_3,jl}(\mathbf{k}_3)
\right]+5~{\rm perms}~{\rm of}~ 1, 2, 3
\right\},\;\;\;
\end{eqnarray}
\end{widetext}
where $K=k_1+k_2+k_3$ and
\begin{eqnarray}
\tilde {\cal A}(k_1,k_2,k_3):=
-\frac{K}{16}\biggl[
1-\frac{1}{K^3}\sum_{i\neq j}k_i^2k_j-4\frac{k_1k_2k_3}{K^3}
\biggr].\;\;\;
\end{eqnarray}
We see that
the second contribution, which is present in the case of Einstein gravity,
is independent of any functions in the Lagrangian,
and hence for all the models of single-field inflation
${\cal A}_{i_1j_1i_2j_2i_3j_3}^{({\rm GR})}$
coincides with the one in general relativity.
(For this reason we associate this piece of the amplitude with
the superscript ``GR.'')
The size of the first contribution,
${\cal A}_{i_1j_1i_2j_2i_3j_3}^{({\rm new})}$, is crucially dependent on
how $\phi$ couples to gravity along the inflationary trajectory.
Only those two amplitudes are sufficient to characterize
the tensor bispectrum in the most general single-field inflation model.

We are now in position to discuss whether or not large
non-Gaussianities can be obtained from ${\cal
A}_{i_1j_1i_2j_2i_3j_3}^{({\rm new})}$.  Since ${\cal G}_T\supset
HX\dot\phi G_{5X}$, the ratio $HX\dot\phi G_{5X}/{\cal G}_T$ cannot be
large in models with large $HX\dot\phi G_{5X}$.  The only possibility is
that various terms in ${\cal G}_T$ are arranged to cancel each other to
give ${\cal G}_T\sim 0$.  (Note, however, that ${\cal G}_T$ must be
finite and positive.)  This then yields ${\cal F}_S\sim -{\cal F}_T$,
where ${\cal F}_S$ is the coefficient of $(\partial\zeta)^2$ in the
quadratic action of the curvature perturbation $\zeta$ and must be
positive to avoid gradient instabilities (see Ref.~\cite{G2}).  Therefore,
models with small ${\cal G}_T$ tend to be unstable against either scalar
or tensor perturbations.
For this reason, generally speaking, it is rather non-trivial to get
large non-Gaussianities from the $\dot h^3$ term, though
one cannot completely deny the possibility of
making both ${\cal F}_S$ and ${\cal F}_T$ positive
with the help of the functional degrees of
freedom of our Lagrangian.

\begin{figure}[t]
  \begin{center}
    \includegraphics[keepaspectratio=true,height=55mm]{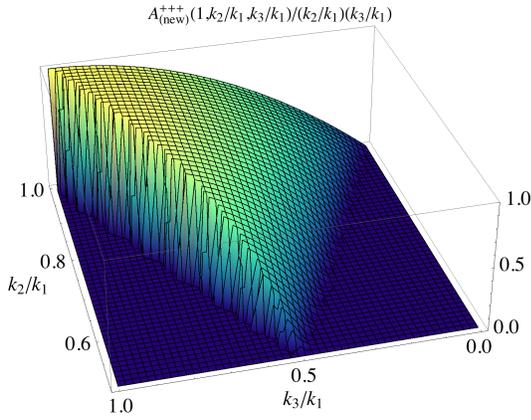}
  \end{center}
  \caption{${\cal A}_{({\rm new})}^{+++}(1, k_2/k_1, k_3/k_1)(k_2/k_1)^{-1}(k_3/k_1)^{-1}$
  as a function of $k_2/k_1$ and $k_3/k_1$. The plot is normalized to unity for
  equilateral configurations $k_2/k_1=k_3/k_1=1$.
  }%
  \label{fig:new-ppp.eps}
\end{figure}

\begin{figure}[t]
  \begin{center}
    \includegraphics[keepaspectratio=true,height=55mm]{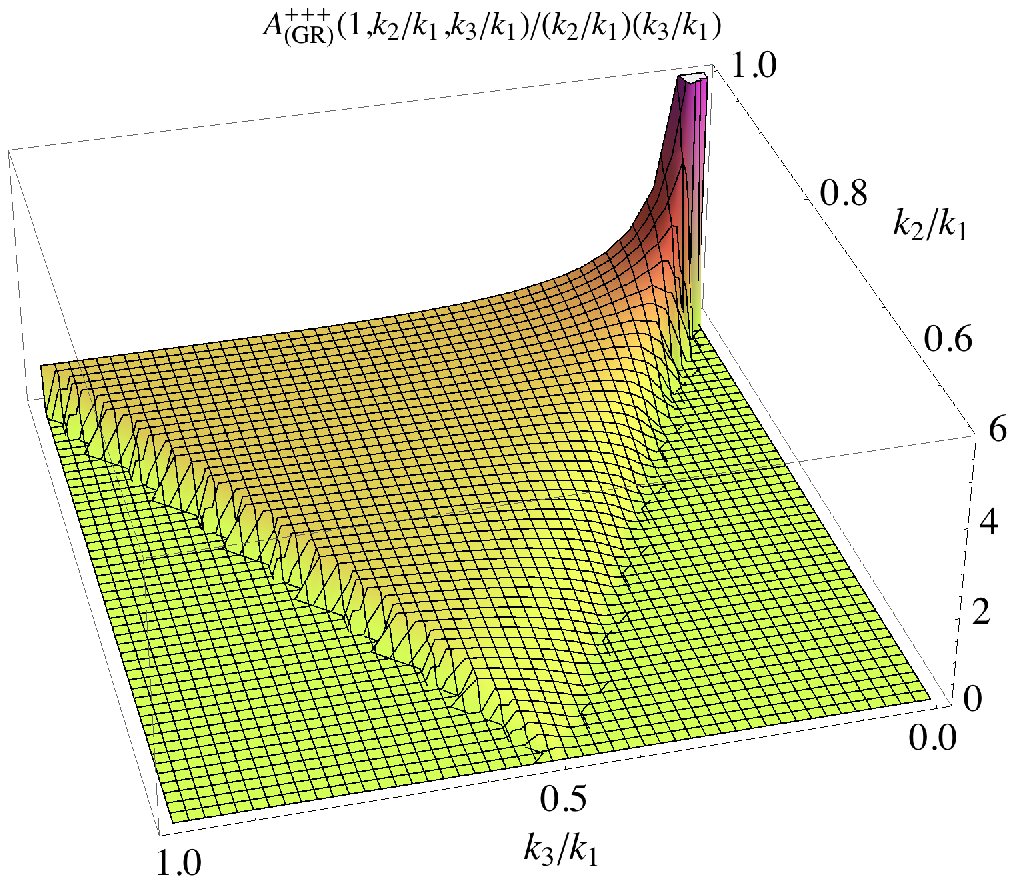}
  \end{center}
  \caption{${\cal A}_{({\rm GR})}^{+++}(1, k_2/k_1, k_3/k_1)(k_2/k_1)^{-1}(k_3/k_1)^{-1}$
  as a function of $k_2/k_1$ and $k_3/k_1$. The plot is normalized to unity for
  equilateral configurations $k_2/k_1=k_3/k_1=1$.
  }%
  \label{fig:GR-ppp.eps}
\end{figure}

Let us turn to the two polarization modes,
\begin{eqnarray}
\xi^{(s)}(\mathbf{k})
:= \tilde h_{ij}(\mathbf{k}) e^{*(s)}_{ij}(\mathbf{k}),
\end{eqnarray}
and consider their amplitudes ${\cal A}^{s_1s_2s_3}$
of the bispectra
$\langle \xi^{s_1}(\mathbf{k}_1)\xi^{s_2}(\mathbf{k}_2)\xi^{s_3}(\mathbf{k}_3)\rangle$.
The amplitude may be defined in an analogous way to Eq.~(\ref{def-amp}),
so that
$
{\cal A}_{({\rm new}),({\rm GR})}^{s_1s_2s_3}=
e_{i_1j_1}^{*(s_1)}(\mathbf{k}_1)
e_{i_2j_2}^{*(s_2)}(\mathbf{k}_2)
e_{i_3j_3}^{*(s_3)}(\mathbf{k}_3)
{\cal A}_{i_1j_1i_2j_2i_3j_3}^{({\rm new}),({\rm GR})}
$.
We thus obtain
\begin{eqnarray}
{\cal A}_{({\rm new})}^{s_1s_2s_3}&=&\frac{HX\dot\phi G_{5X}}{4{\cal G}_T}
\frac{k_1^2k_2^2k_3^2}{K^3}F_{({\rm new})}^{s_1s_2s_3}(k_1,k_2,k_3),
\\
{\cal A}_{({\rm GR})}^{s_1s_2s_3}&=&\tilde{\cal A}(k_1,k_2,k_3)
F_{({\rm GR})}^{s_1s_2s_3}(k_1,k_2,k_3),
\end{eqnarray}
where we defined
\begin{eqnarray}
&&F_{({\rm new})}^{+++}(k_1, k_2, k_3):=
\frac{K^3}{64k_1^2 k_2^2k_3^2}
\nonumber\\&&\qquad\qquad\qquad\qquad \times
\biggl[K^3-4
\sum_{i\neq j}k_i^2k_j-4 k_1k_2k_3 \biggr],\;\;\;\;\;\;\;\;
\end{eqnarray}
$F_{({\rm GR})}^{+++}(k_1,k_2,k_3):=K^2 F_{({\rm new})}^{+++}(k_1,k_2,k_3)/2$,
and $F_{({\rm new}),({\rm GR})}^{++-}(k_1,k_2,k_3):=F_{({\rm new}),({\rm GR})}^{+++}(k_1,k_2,-k_3)$.
Since our theory accommodates no parity violation,
we have $F^{---}_{({\rm new}),({\rm GR})}=F^{+++}_{({\rm new}),({\rm GR})}$
and
$F^{--+}_{({\rm new}),({\rm GR})}=F^{++-}_{({\rm new}),({\rm GR})}$.

The non-Gaussian amplitudes ${\cal A}_{({\rm new})}^{+++}$ and ${\cal
A}_{({\rm GR})}^{+++}$ are plotted in Figs.~\ref{fig:new-ppp.eps}
and~\ref{fig:GR-ppp.eps}.  One sees that the amplitude of the new
contribution peaks in the equilateral configuration, while the ``GR''
contribution becomes largest in the squeezed limit.
This gives a clear distinction between the two different contributions, and
the two characteristic shapes would be
helpful to discriminate the inflationary gravitational waves from those produced
by other sources. The other correlation functions such as the $-++$ one
are subdominant relative to the $+++$ one because for equilateral
configurations $F_{({\rm new})}^{++-}=F_{({\rm new})}^{+-+}=F_{({\rm
new})}^{-++}=F_{({\rm new})}^{+++}/9$ and $F_{({\rm GR})}^{++-}=F_{({\rm
GR})}^{+-+}=F_{({\rm GR})}^{-++}=F_{({\rm GR})}^{+++}/81$, and in the
squeezed limit, $\mathbf{k}_3\to 0$, one has
$F_{({\rm GR})}^{+++}\approx F_{({\rm GR})}^{++-}\approx -k_1^2/2$
and
$F_{({\rm GR})}^{+-+}\approx F_{({\rm GR})}^{-++}\approx -k_3^4/32k_1^2$.

In this {\em Letter},
we have clarified
primordial non-Gaussianities of
tensor perturbations arising from the most
general single-field inflation model with second-order field
equations, and
have found that they are
completely determined by two different contributions: ${\cal A}_{({\rm new})}$ and ${\cal A}_{({\rm GR})}$.
Our results provide at least two distinctive features to
test the framework of generalized G-inflation based on the graviton
non-Gaussianities.  Firstly, ${\cal A}_{({\rm new})}$ is a {\em unique}
feature of the kinetic coupling term $G_5$.  Any detection of this type
of bispectrum, no matter large or small, would unambiguously indicate
the existence of non-vanishing $G_{5X}$, at least in the Galileon
framework.  Secondly, the contribution ${\cal A}_{({\rm GR})}$ is a {\em
fixed} and {\em universal} feature for single-field inflation models
which are all within the generalized G-inflation framework.  It is
impossible to enhance/suppress this contribution in generalized
G-inflation models. In other words, any detection of
enhancement/suppression of this contribution to the graviton bispectrum
would imply new physics beyond generalized G-inflation and/or other
astrophysical sources.  The two contributions are clearly
distinguishable according to their shapes of non-Gaussianities.

\paragraph*{Acknowledgments}
This work was
supported in part by
ANR (Agence Nationale de la Recherche)
grant ``STR-COSMO'' ANR-09-BLAN-0157-01 (X.G.),
JSPS Grant-in-Aid for Research Activity Start-up
No.~22840011 (T.K.), the Grant-in-Aid for Scientific Research
Nos.~21740187 (M.Y.), 23340058 (J.Y.), and the Grant-in-Aid for
Scientific Research on Innovative Areas No.~21111006 (J.Y.).


\end{document}